\documentclass[reprint, aps, prx, floatfix,superscriptaddress]{revtex4-2}

\usepackage{graphicx}    
\usepackage{dcolumn}    
\usepackage{bm}             
\usepackage{amssymb}   
\usepackage{amsmath}    
\usepackage{amsthm}
\usepackage{mathrsfs}    
\usepackage{commath}   
\usepackage{subfigure}    
\usepackage{braket}
\usepackage{color}
\usepackage{siunitx}
\usepackage[colorlinks=true, allcolors=blue]{hyperref}
\usepackage{hyphenat}
\usepackage{footnote}
\usepackage{notes2bib}
\usepackage{xcolor}

\newcommand{\beq}{\begin{equation}}
\newcommand{\eeq}{\end{equation}}

\begin{document}

\title{Efficiently catching entangled microwave photons from a quantum transducer with shaped optical pumps}

\author{Changchun Zhong}
\email{zhong.changchun@xjtu.edu.cn}

\affiliation{Department of Physics, Xi'an Jiaotong University, Xi'an, Shanxi 710049, China}
\affiliation{SeQure, Chicago, IL, 60615, USA}

\date{\today}

\begin{abstract}    
Quantum transducer, when working as a microwave and optical entanglement generator, provides a practical way of coherently connecting optical communication channels and microwave quantum processors. The recent experiments on quantum transducer verifying entanglement between microwave and optical photons show the promise of approaching that goal. While flying optical photons can be efficiently controlled or detected, the microwave photon needs to be stored in a cavity or converted to the excitation of superconducting qubit for further quantum operations. However, to efficiently capture or detect a single microwave photon with arbitrary time profile remains challenging. This work focuses on this challenge in the setting of entanglement-based quantum transducer and proposes a solution by shaping the optical pump pulse. By Schmidt decomposing the output entangled state, we show the microwave-optical photon pair takes a specific temporal profile that is controlled by the optical pump. The microwave photon from the transducer can be absorbed near perfectly by a receiving cavity with tunable coupling and is ready to be converted to the excitation of superconducting qubits, enabling further quantum operations.
\end{abstract}

\maketitle

\textit{Introduction}--Quantum communication with optical photons and quantum computation based on superconducting circuits form the two major modules of the modern fast-developing quantum technology. The coherent combination of the two fields is essential for the long-pursued goal of quantum network and the design of modular quantum architecture \cite{cirac1997,kimble2008}. However, due to the large energy difference, optical and microwave photons do not naturally interact, thus coherently converting quantum information between them is extremely challenging given the state-of-the-art technology \cite{lauk2020,avossa2024}. 

Among all recent quantum transduction models, entanglement-based quantum transducer (EQT) shows the experimental feasibility \cite{zhong2020}. EQT relies on physical platforms that first generate microwave-optical entanglement, which is further used as a quantum resource for microwave-optical quantum information teleportation \cite{zhong2022PRAPP,wu2021} or connecting microwave quantum circuits through the well-known DLCZ protocol \cite{duan2001,zhong2020A,stefan2021}. The recent experiments verifying the microwave-optical entanglement take an important step toward EQT \cite{sahu2023,meesala2023,meesala2023B}. {It is worth noting that verifying entanglement generally destroys the entangled states \cite{van2007}. For practical application, we need to either store the entangled state or perform fast quantum operations (including detection) on the photons. While the detection of single optical photon is well-developed, its counterpart in microwave frequency is less-advanced due to its low energy scale. The circuit QED system offers a promising approach to accomplishing microwave photon control where the microwave photon is first absorbed by a receiving cavity, enabling non-demolish photon detection \cite{kono2018,royer2018,besse2018}. Moreover, the microwave photon can later converted to qubit excitation for further quantum operations, e.g., quantum gates and quantum measurements \cite{blais2021}.}

A general receiving cavity can only perfectly capture a microwave with an exponentially growing temporal profile \cite{wenner20144}. Unfortunately, due to the finite life time, a photon generated from most quantum systems naturally takes an exponentially decaying time profile. Capturing such a photon using a receiving cavity practically can only achieve finite efficiency even when the cavity has a certain coupling rate tunability. {A well-known way to improve the efficiency is to make the releasing cavity tunable so as to generate a microwave photon with exponential increasing time profile, known as "pitch and catch" protocol \cite{campagne2018,wenner20144,axline2018,asaf2023}.} However, such kind of realization imposes practical limits on the experiment design, especially quantum transducers with on-chip structure \cite{meesala2023,han2020}. 

Here, we show that, by using a controlled optical pump pulse, a general quantum transducer can generate entangled microwave-optical photon pairs such that the microwave photon carries a temporal profile which can be absorbed perfectly by a receiving cavity with practical tunability. It is achievable mainly due to the transducer's structure: a quantum transducer is composed of coupled microwave and optical modes with the coupling strength determined by the pump photons. If the pump is controlled to be first smoothly increasing then decreasing, the probability of photon generation will approximately follow the same shape, which indicates the single photon would carry a temporal profile that smoothly goes up and down. Such kind of photon can be captured \textit{in principle} with almost $100$\% efficiency if we properly tune the coupling strength of the receiving cavity in such a way that the reflection of the incoming photon and the leakage of the receiving cavity destructively interfere.



\begin{figure}[t]
\centering
\includegraphics[width=\columnwidth]{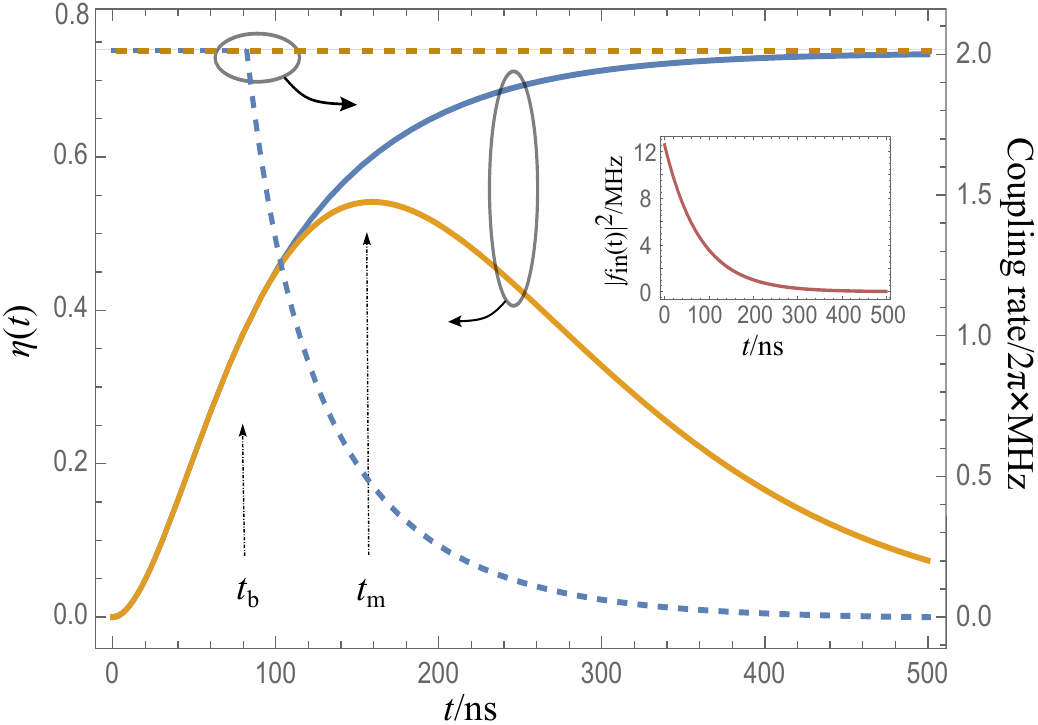}
\caption{The capture efficiency (solid curves) and the cavity coupling strength (dashed lines) in terms of time. The solid orange curve shows the efficiency (maximal achieved at $t_m$) for fixing cavity coupling rate at $\kappa_1=2\pi\times2.0$ MHz (indicated by the dashed orange line). The blue solid curve gives the efficiency with a tunable cavity coupling. The blue dashed line shows how to tune $\kappa_1(t)$ for $t>t_b$. The input microwave photon is assumed to have a natural decay tail with decay rate $\gamma=2\pi\times2.0$ MHz, as shown by the red curve in the inset. {The ovals encircle the curves of capture efficiencies and coupling rates, as indicated by the arrows.} \label{fig1}}
\end{figure}

\textit{A receiving cavity with tunable coupling rate}--We first present the theory of catching a single microwave photon that takes a naturally exponential decaying temporal profile $f_\text{in}(t)=\sqrt{\gamma}e^{-\frac{\gamma}{2} t}$, where $\gamma$ is the energy decay rate. A receiving cavity with mode operator $\hat{d}$ is used to catch this microwave photon. The question to ask is how efficient can the photon be absorbed by the cavity. To answer that, we check the dynamics of the cavity mode which is governed by $\dot{d}(t)=-\frac{\kappa_1}{2}d(t)+\sqrt{\kappa_1}f_\text{in}(t)$, where $\kappa_1$ is the cavity coupling rate. For the moment, we keep $\kappa_1$ constant and later we would allow it to be tunable. The cavity mode amplitude is solved as
\begin{equation}
    d(t)=\frac{2\sqrt{\gamma\kappa_1}}{\kappa_1-\gamma}(e^{-\frac{1}{2}\gamma t}-e^{-\frac{1}{2}\kappa_1t}).
\end{equation}
If we define an energy capture efficiency as $\eta(t)\equiv\abs{d(t)}^2/\int_0^\infty \abs{f_\text{in}(t)}^2dt$, we have
\begin{equation}
    \eta(t)=\frac{4\gamma\kappa_1(e^{-\frac{1}{2}\gamma t}-e^{-\frac{1}{2}\kappa_1t})^2}{(\kappa_1-\gamma)^2}.
\end{equation}
It is straightforward to show that $\eta(t)$ achieves the maximal value at $t_m\equiv\frac{2}{\kappa_1-\gamma}\ln\frac{\kappa_1}{\gamma}$, which corresponds to the time when the cavity instantaneous leakage becomes larger than the energy being captured. Since the cavity leakage exists at all time, the photon capture efficiency is always less than one. In fact, the maximal efficiency one can achieve is $\eta_\text{max}=4/e^2\simeq 54\%$ by choosing $\kappa_1\sim\gamma$, as shown by the orange curve in Fig.~\ref{fig1}, which is far from satisfaction.

One way to improve the efficiency is to make the coupling rate $\kappa_1$ tunable. Imagine at some point that the cavity leakage and the input photon reflection destructively interfere. In the frame work of input-output theory, it satisfies
\begin{equation}
    d_\text{out}(t_b)=f_\text{in}(t_b)-\sqrt{\kappa_1}d(t_b)=0.
\end{equation}
We will define this as \textit{the balance condition} throughout the paper and $t_b$ as \textit{the balance time}. It is straightforward to show $t_b=\frac{2}{\kappa_1-\gamma}\ln{\frac{2\kappa_1}{\kappa_1+\gamma}}$ in this case. It turns out that this balance condition can also be fulfilled at any later time ($t>t_b$) if we can properly tune down the coupling rate. Effectively, all energy beyond the time $t_b$ will be stored inside the cavity, thus increasing the final capture efficiency. To realize that, we need to tune the coupling rate $\kappa_1$ in a way such that 
\begin{equation}\label{tdc}
    \sqrt{\kappa_1(t)}d(t)=f_\text{in}(t)
\end{equation}
is satisfied for all time $t\ge t_b$. To determine $\kappa_1(t)$ for $t>t_b$, we first solve the cavity mode as $d(t)=(\abs{d(t_b)}^2+\int_{t_b}^t\abs{f_\text{in}(\tau)}^2d\tau)^{1/2}$ for $t>t_b.$
This equation can be understood as that all later incoming energy is accumulated in the cavity. Using the expression of $t_b$ and perform the integral, we have
\begin{equation}
    d(t)=\left({e^{-\gamma t_b}-e^{-\gamma t}+\frac{4\gamma\kappa_1({e^{-\frac{1}{2}\gamma t_b}-e^{-\frac{1}{2}\kappa_1 t_b} })^2}{(\kappa_1-\gamma)^2} }\right)^{1/2}.
\end{equation}
We can similarly define an energy efficiency whose maximal is achieved at infinite time
\begin{equation}\label{efc2}
    \eta(\infty)=e^{-\gamma t_b}+\frac{4\gamma\kappa_1(e^{-\frac{1}{2}\gamma t_b}-e^{-\frac{1}{2}\kappa_1t_b})^2}{(\kappa_1-\gamma)^2}.
\end{equation}
Using Eq.~\ref{tdc}, the expression $\kappa_1(t)$ can be obtained 
\begin{equation}
    \kappa_1(t)=\frac{\kappa_1\gamma e^{-\gamma t}}{\kappa_1(e^{-\gamma t_b}-e^{-\gamma t})+\gamma e^{-\gamma t_b}}, \text{ for }t\in[t_b,\infty],
\end{equation}
which satisfies the continuous condition $\kappa_1(t_b)=\kappa_1$. The efficiency Eq.~\ref{efc2} can in principle achieve unit value if initially we have an infinite large coupling rate $\kappa_1$. However, for quantum transducers, $\kappa_1$ is finite practically and is usually on the order of $\gamma$, which makes the achievable efficiency around $80\%$, as shown by the blue curve in Fig.~\ref{fig1}.

\textit{Catching microwave photons with unit efficiency}--If one can reverse the temporal profile of the input photons, $100\%$ capture efficiency is possible \cite{campagne2018,asaf2023,wenner20144}. This requires the input photon has an exponential increasing temporal profile $f_\text{in}(t)=\sqrt{\gamma}e^{\gamma t/2}$ for $t\in(-\infty,0]$. Intuitively, the perfect absorption can be understood as the reverse process of a cavity naturally emitting a photon. It is more obvious when solving the cavity mode to be $d(t)=e^{\gamma t/2}$, where the cavity coupling rate is chosen as $\kappa_1=\gamma$. We see the destructive interference condition can be satisfied during the whole capture process $\sqrt{\gamma}d(t)=f_\text{in}(t)$, which means all input energy will accumulate inside the cavity.

This suggests that $100\%$ capture efficiency is possible for photons with more general temporal profile. For example, a microwave photon with the following time profile
\begin{equation}\label{egtd}
f_\text{in}(t)=
\begin{cases} \sqrt{\frac{\gamma_1\gamma_2}{\gamma_1+\gamma_2}}e^{\frac{\gamma_1}{2}(t-t_0)} ,&t\le t_0  \\
      \sqrt{\frac{\gamma_1\gamma_2}{\gamma_1+\gamma_2}}e^{-\frac{\gamma_2}{2}(t-t_0)},& t> t_0
\end{cases},     
\end{equation}
which initially grows with rate $\gamma_1$ until time $t_0$, then decays with rate $\gamma_2$. Following a similar approach, one can show that such a microwave photon can be absorbed with unit efficiency by a cavity with the coupling rate tuned in the following way
\begin{equation}
    \kappa_1(t)=\begin{cases}
        \gamma_1, &t\le t_0 \\
        \frac{\gamma_1\gamma_2}{(\gamma_1+\gamma_2)e^{\gamma_2(t-t_0)}-\gamma_1}, &t>t_0
    \end{cases}.
\end{equation}
The energy capture efficiency grows as follows
\begin{equation}
    \eta(t)=\begin{cases}
        \frac{\gamma_2}{\gamma_1+\gamma_2} e^{\gamma_1(t-t_0)},&t\le t_0\\
        \frac{\gamma_2}{\gamma_1+\gamma_2}+\frac{\gamma_1}{\gamma_1+\gamma_2}(1-e^{-\gamma_2(t-t_0)}),&t>t_0
    \end{cases}.
\end{equation}
The analytical expressions clearly shows that, by tunning properly the cavity coupling strength, the efficiency could reach $100\%$ as time goes to infinity. 

Admittedly, it is not always possible to get a microwave photon that carries the temporal profile as in Eq.~\ref{egtd}. In practice, the photons could have more general temporal profiles. In order to maximize the capture efficiency, it turns out that the key is to minimize the initial reflection before the cavity leakage destructively interferes with all further reflection (balanced). This can be achieved if we can control the input photons to have growing temporal profiles, such that the balance condition is satisfied in a very early time while most of the photon energy is still yet to be captured. In the following, we show that the microwave photon from a quantum transducer can be controlled in such a specific temporal form, and it can be absorbed almost perfectly by a receiving cavity with properly tuned coupling strength.

\begin{table*}[t]
\caption{The following experimental feasible parameters for a cavity electro-optic system are used in the numerical evaluations in the text (unless specified otherwise). {$\kappa_\text{e,i}$ and $\kappa_\text{e,c}$ ($\kappa_\text{o,i}$ and $\kappa_\text{o,e}$) refer to the electrical (optical) mode intrinsic and external coupling rate, respectively.} } 
\label{tab1}
\begin{center}
\begin{tabular}{c|c|c|c|c|c|c}
\hline
\hline
$\kappa_ {\text{e,i}}$/MHz & $\kappa_ {\text{e,c}}$/MHz & $\kappa_{\text{o,i}}$/(GHz)   & $\kappa_{\text{o,c}}$/(GHz)  & $g_0$/kHz & $\omega_o$/THz &  $\omega_e$/GHz \\
\hline
$2\pi\times 0.55$   & $2\pi\times 1.25$ &  $2\pi\times 0.65$   & $2\pi\times 0.65$  &   $2\pi\times 260$ & $2\pi\times 190$ & $2\pi\times 5$    \\
\hline
\hline
\end{tabular}
\end{center}
\end{table*}

\begin{figure*}[t]
\centering
\includegraphics[width=\textwidth]{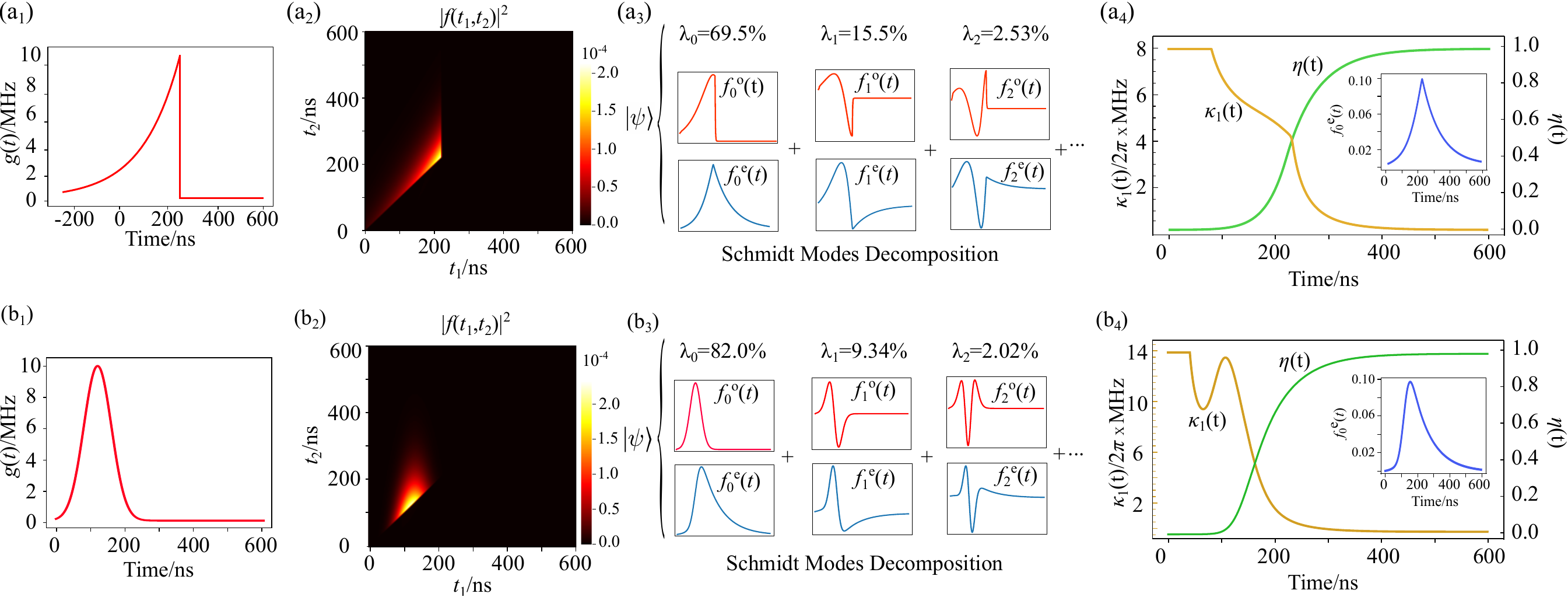}
\caption{Entangled photon pair generation and wave packet analysis for two different model squeezing strength (upper and lower rows). {The first row shows the case with a piecewise squeezing strength that initially grows exponentially, while the second row gives the case with a Gaussian squeezing temporal profile. The four columns (from left to right) show the squeezing pulse shapes($a_1,b_1$), output wave packets ($a_2,b_2$), Schmidt decomposed temporal modes ($a_3,b_3$) and the capture efficiency while tuning the coupling strength ($a_4,b_4$), respectively. The blue curves in the insets of ($a_4,b_4$) highlight the temporal shape of the incoming microwave photons.} \label{fig2}}
\end{figure*}



\textit{A model system for quantum transducer}--Quantum transducer is studied actively based on various physical platforms which generally feature certain nonlinear coupling between the microwave and optical modes \cite{lauk2020,han2021}. \textit{Without loss of generality}, we consider a cavity electro-optic system for illustration, which involves a three-wave mixing interaction among the optical cavity modes and a microwave superconducting resonator enabled by certain material with Pockels nonlinearity \cite{fan2018}. Using a blue detuned optical drive, the three wave mixing interaction usually reduces to a two-mode-squeezing coupling {with rotating wave approximation}, in which case the quantum transducer is sometimes called optical-microwave entanglement generator.

We denote $\hat{a}$ and $\hat{c}$ as the optical and microwave mode operators, 
$\omega_o$ and $\omega_e$ as the corresponding mode frequencies, respectively. {The optical pump is detuned in the blue side band, which satisfies $\omega_p=\omega_o+\omega_e$, as shown in the pink panel of Fig.~\ref{fig2}.} The Hamiltonian is expressed as follows \cite{holzgrafe2020}
\begin{equation}\label{hamil}
\begin{split}
\hat{H}/\hbar=& \omega_o\hat{a}^\dagger\hat{a}+\omega_e\hat{c}^\dagger\hat{c}-g_0\sqrt{\bar{n}_o}(\hat{a}^\dagger\hat{c}^\dagger+\hat{a}\hat{c}).
\end{split}
\end{equation}
Here $g_0$ represents the one-photon electro-optic coupling rate. The coupling can be further enhanced by the intracavity photon number $\bar{n}_o$ induced by the laser pump. We define $g(t):= g_0\sqrt{\bar{n}_o}$ as the squeezing strength. We consider a shaped pump pulse that induces a time dependent intracavity photon $\bar{n}_o(t)$. This means that the squeezing strength would take a specific form in time, which could control the temporal profile of the output photons from the transducer. In this paper, we take the feasible parameters for numerical discussions as listed in Tab.~\ref{tab1}. More detailed descriptions of electro-optic system can be found in Ref.~\cite{tsang2010,tsang2011,fan2018,youssef2022}.

\textit{The bi-photon wave packet and the Schmidt temporal modes}--A quantum transducer with a weak-blue-detuned laser pump is able to generate entangled optical-microwave photon pairs with the help of the induced two-mode squeezing interaction \cite{vitali2012}. The pair-photon generated from this process can be approximately described by the so-called bi-photon wave packet $\ket{\psi}=\iint dt_1dt_2f(t_1,t_2)\hat{a}^\dagger(t_1)\hat{c}^\dagger(t_2)\ket{0}$ \cite{brecht2015}. The coefficient $f(t_1,t_2)$ is related to the two-time correlation function \cite{glauber1963}
\begin{equation}\label{eq_co}   \abs{f(t_1,t_2)}^2\propto
\begin{cases} \braket{\hat{a}^\dagger(t_1)\hat{c}^\dagger(t_2)\hat{c}(t_2)\hat{a}(t_1)}, &t_1< t_2  \\    \braket{\hat{c}^\dagger(t_2)\hat{a}^\dagger(t_1)\hat{a}(t_1)\hat{c}(t_2)},&t_2< t_1
\end{cases},    
\end{equation}
which is understood as the probability of getting an optical photon at time $t_1$ and a microwave photon at time $t_2$. In order to find the photon temporal mode, we decompose the bi-photon wave packet into the temporal orthogonal modes through the well-known Schmidt decomposition $f(t_1,t_2)=\sum^\infty_{k=0}\sqrt{\lambda_k}f^o_k(t_1)f^e_k(t_2)$ \cite{watrous2018,kok2010}. The wave packet can be rewritten as 
\begin{equation}    \ket{\psi}=\sum_{k=0}^\infty\sqrt{\lambda_k}\ket{\psi_k^o}\ket{\psi_k^e},
\end{equation}
where the states $\ket{\psi_k^o}=\int dt_1f^o_k(t_1)\hat{a}^\dagger(t_1)\ket{0}$ and $\ket{\psi_k^{e}}=\int dt_2 f^e_k(t_2)\hat{c}^\dagger(t_2)\ket{0}$ are the optical and microwave $k_\text{th}$ temporal mode, respectively. We see pairs of optical and microwave temporal modes are excited with probability $\lambda_k$. The state is entangled in the temporal mode degrees of freedom with the entanglement entropy $S=-\sum_k\lambda_k\ln\lambda_k$. In principle, the temporal modes span an infinite dimensional space, and efficiently controlling them has a great potential for quantum information processing \cite{brecht2015}.

\textit{Controlling the microwave temporal mode}--We emphasize that in practice, a versatile laser pulse can be generated \cite{monmayrant2010}, which could be used to control the squeezing strength effectively and thus shape the output photon temporal profile. For demonstration, we consider two different models: using shaped laser pump pulses (with the duration of several hundred nanoseconds) that generate the following two different time dependent squeezing strength
\begin{equation}
\begin{split}
    g(t)&= \begin{cases}
        \mathcal{G}_1 e^{\frac{\gamma}{2}(t-\mu)}*g_0, &t<\mu\\
        0, &t\ge \mu        
    \end{cases},\\
    g(t)&= \mathcal{G}_2 e^{-\frac{(t-\nu)^2}{2\sigma^2}}*g_0,\\
\end{split}    
\end{equation}
The first expression is a piecewise function which grows exponentially in the very beginning and takes zero value after time $\mu$. The second one is a standard Gaussian function. As an example for numerical evaluation, we fixing the parameters $\mathcal{G}_1=5.5$, $\gamma=12$ MHz, $\mathcal{G}_2=6.5$, $\sigma=40$ ns and $(\mu,\nu)=(220\text{ ns},120\text{ ns})$ (the two time dependent squeezing strength are schematically shown in Fig.~\ref{fig2}(a$_1$) and \ref{fig2}(b$_1$)). We numerically solve the time dependent Hamiltonian Eq.~\ref{hamil} with the help of QuTip Python package \cite{qutip2012}. In the calculation, all modes are assumed to couple to vacuum baths since: 1) the optical mode frequency is several hundreds THz, leading to almost zero thermal photon even at room temperature; 2) the transducer is placed in dilute refrigerator with temperature on the order of several mK, yielding a negligible thermal noise for the microwave modes with several GHz mode frequency. Note the laser pump might induce unwanted device heating, which can be included by replacing the vacuum bath with certain thermal noise. {To catch the main idea, in this paper we assume the laser pump pulse is small and short enough such that the above conditions are always true.}

In order to get the bi-photon wave packet, we first calculate the two time correlation function Eq.~\ref{eq_co}, which can be obtained with the help of quantum regression theorem \cite{gardiner2004}. As shown in Fig.~\ref{fig2}(a$_2$), in the first $220$ ns, the microwave photon is probable to be detected after the optical photon click. This is due to the relatively small microwave coupling rate $\kappa_e\ll\kappa_o$. Beyond $220$ ns, the probability for coincident counting the photon pairs sharply drops since the squeezing strength drops as used in this model. Similarly, the Fig.~\ref{fig2}(b$_2$) shows the output correlation function when the squeezing strength takes a Gaussian time profile. The blob on the upper left region indicates the non-zero chance of detecting a microwave photon after the optical photon is already detected. 

To obtain the microwave temporal profiles, we further perform the Schmidt decomposition of the wave packets, and the results are shown in Fig.~\ref{fig2}(a$_3$) and \ref{fig2}(b$_3$). We see that the Schmidt zero mode has the largest probability and it is the mode that we are interested in. As shown in Fig.~\ref{fig2}(a$_3$) and \ref{fig2}(b$_3$), the optical photon zero modes have temporal profiles that closely follow the squeezing strength, resulting from the large optical cavity coupling rate. Meanwhile, the microwave photon zero modes have obvious decaying tails since the photon slowly couples out of the system with a smaller coupling rate. More importantly, the temporal profiles of microwave photons gradually increase in time for both models (not necessarily exponentially growing), suggesting the possibility of capturing the photons with high efficiency.

Figure~\ref{fig2}(a$_4$) and \ref{fig2}(b$_4$) give the numerical results. As shown by the green curves, the capture efficiency gradually approaches the unity while we tune the receiving cavity coupling rate. The way we tune the coupling rate is delineated by the orange lines. For the first case, the microwave photon carries a time profile (depicted in the inset of Fig.~\ref{fig2}(a$_4$)) that closely resembles the ideal model in Eq.~\ref{egtd}. Note for the ideal model Eq.~\ref{egtd}, the balance condition can always be satisfied, which means all energy will in principle be captured. For current more practical case, we similarly keep a constant $\kappa_1$ in the beginning and wait until the balance condition is achieved, then we start tuning down the coupling rate to make sure the balance condition is fulfilled at all later time. Since the balance condition is not satisfied in the beginning, some photon energy is lost which reduces the total capture efficiency. Luckily, the energy loss is small due to: 1) most of the photon energy is coming in later time; 2) the balance condition is nearly satisfied. The above analysis is similarly true for the second case. It is worth noting that the way we tune the coupling rate differs. As shown by the orange curve in Fig.~\ref{fig2}(b$_4$), there is an obvious dip, resulting from the balance condition which requires a lower leakage to cancel the initially small incoming reflection. This indicates a more decent tuning of the cavity coupling rate might be needed for general pump pulses.

\textit{Discussion}--{The laser pump has long been used as an extra degree of freedom to control photon temporal modes \cite{raymer2020,arzani2018}, e.g., in the setting of quantum optics it was used to phase match the parametrically down converted photons with atomic transitions \cite{carnio2021}. In this paper, as a cutting-edge technique for quantum transducer, we analyzed the cavity-capturing efficiency of microwave photons generated from electro-optics with two typical model squeezing strength (controlling the laser pump).} Actually, the method could be more general in the sense of: 1) being applied to other physical platform based quantum transducers, such as electro-optomechanics; 2) using some squeezing strength with more general time profile. 

The model squeezing strength we discussed may not be the optimal choice. The energy capturing efficiency can approach almost unit value as long as the microwave photon carries a time profile that is beneficial to achieving the balance condition as early as possible and very little energy is lost in the meantime. The unit capturing efficiency might be obtainable: 1) tuning squeezing strength with different time profiles which generate optimal shaped microwave photons; 2) using cavity with flexible tunable coupling rates. Certain choices of laser pump may also lead to a large occupation of the photons in the Schmidt zeroth mode, which is eventually helpful in increasing the probability of catching the entangled photons from the transducers. A very promising method to find the optimal pump is to first fix the output Schmidt mode profiles where the microwave photon can be captured perfectly. Then the squeezing strength or the pump pulse is to be derived through \textit{reverse engineering}, and we leave this interesting topic for future study.

In practice, how we choose the parameters depends on the detailed experimental setup. If we have more flexibility in controlling the squeezing strength, the requirement of cavity coupling rate tunability could be designed to be less demanding. Or if we are more limited by the control of squeezing, we might need to devote more to the design of receiving cavity. In either case, the parameters shall be optimized in order to achieve high enough capturing efficiency. Our method provides a systematic way of doing that, which is not only an important step in finally achieving the ambitious quantum networks but also will be beneficial to other applications related to microwave quantum engineering and quantum operations.

\begin{acknowledgments}
C.Z. thanks Mankei Tsang and Liang Jiang for helpful discussions. C.Z. thanks the start up support from Xi'an Jiaotong University (Grant No. 11301224010717).
\end{acknowledgments}

\bibliography{all}

\onecolumngrid

\begin{appendix}

\end{appendix}

\end{document}